\documentstyle[aaspp4]{article}

\begin{document}

\title{Spectra and Light Curves of Gamma-Ray Burst Afterglows} 
\author{Re'em Sari, Tsvi Piran} 
\affil{Racah Institute, Hebrew University, Jerusalem 91904, Israel }
\authoremail{sari@nikki.fiz.huji.ac.il,tsvi@nikki.fiz.huji.ac.il} 
\and
\author{Ramesh Narayan} 
\affil{Harvard-Smithsonian Center for Astrophysics, 60 Garden Street,
Cambridge MA 02138, U.S.A. } 
\authoremail{narayan@surya.harvard.edu}

\begin{abstract}
  The recently discovered GRB afterglow is believed to be described
  reasonably well by synchrotron emission from a slowing down
  relativistic shell that collides with an external medium. To compare
  theoretical models with afterglow observations we calculate here the
  broad band spectrum and corresponding light curve of synchrotron
  radiation from a power-law distribution of electrons in an expanding
  relativistic shock. Both the spectrum and the light curve consist of
  several power-law segments.  The light curve is constructed under
  two limiting models for the hydrodynamical evolution of the shock:
  fully adiabatic and fully radiative. We compare the results with
  observations of $\gamma$-ray burst afterglows.
\end{abstract}


\section{Introduction}
Delayed emission in X-ray, optical and radio wavelengths has been
recently seen in a few $\gamma$-ray bursts (Costa et al. 1997, Groot
et al. 1997, Frail et. al. 1997). This so-called ``afterglow'' is
described reasonably well as synchrotron emission from accelerated
electrons when a spherical relativistic shell collides with an
external medium (Waxman 1997a,b, Wijers, Rees \& M\'esz\'aros 1997,
Katz \& Piran 1997). Previous analyses have described the spectrum and
light curve only over a limited range of frequency and time. In this
Letter we discuss the spectrum over a wide range of frequency and
derive the shape of the light curve from very early to late times. We
focus on the optical and X-ray emission, where synchrotron self
absorption is not important, and we assume that the shell is ultra
relativistic. We allow for both adiabatic and radiative hydrodynamical
evolution.

Consider a relativistic shock propagating through a uniform cold
medium with particle density $n$. Behind the shock, the particle
density and the energy density are given by $4\gamma n$ and $4
\gamma^2 n m_p c^2$, respectively, where $\gamma$ is the Lorentz
factor of the shocked fluid (Blandford \& McKee 1976). We assume that
electrons are accelerated in the shock to a power law distribution of
Lorentz factor $\gamma_e$, with a minimum Lorentz factor $\gamma_m$:
$N(\gamma_e)d\gamma_e\propto \gamma_e^{-p}d\gamma_e$,
$\gamma_e\geq\gamma_m$. To keep the energy of the electrons finite we
take $p>2$. We assume that a constant fraction $\epsilon_e$ of the
shock energy goes into the electrons. Then
\begin{equation}
\gamma_m=\epsilon_e\left(\frac {p-2} {p-1}\right) \frac {m_p} {m_e}
 \gamma \cong 610 \epsilon_e \gamma,
\end{equation}
where the coefficient on the right corresponds to the standard choice,
$p=2.5$ (Sari, Narayan \& Piran 1996). We also assume that the magnetic
energy density behind the shock is a constant fraction $\epsilon_B$ of
the shock energy. This gives a magnetic field strength
\begin{equation}
B=(32 \pi m_p\epsilon_Bn)^{1/2}\gamma c.
\end{equation}

In this Letter we consider only synchrotron emission, and neglect
inverse Compton scattering (which can be important when
$\epsilon_B>\epsilon_e$, Sari et al. 1996).

\section{Synchrotron Spectrum of a Relativistic Shock}
A relativistic electron with Lorentz factor $\gamma_e \gg 1$ in a
magnetic field $B$ emits synchrotron radiation. The radiation power
and the characteristic frequency are given by (Rybicki \& Lightman
1976)
\begin{equation}
\label{power}
P(\gamma_e)=\frac 4 3 \sigma_T c \gamma^2 \gamma_e^2 \frac {B^2} {8\pi},
\end{equation}
\begin{equation}
\label{freq}
\nu(\gamma_e)=\gamma \gamma_e^2 \frac {q_e B} {2 \pi m_e c},
\end{equation}
where the factors of $\gamma^2$ and $\gamma$ are introduced to
transform the results from the frame of the shocked fluid to the frame
of the observer. The spectral power, $P_\nu$ (power per unit
frequency, ${\rm erg\,Hz^{-1}\,s^{-1}}$), varies as $\nu^{1/3}$ for
$\nu<\nu(\gamma_e)$, and cuts off exponentially for
$\nu>\nu(\gamma_e)$ (Rybicki \& Lightman 1976). The peak power occurs
at $\nu(\gamma_e)$, where it has the approximate value
\begin{equation}
\label{flux}
P_{\nu,max}\approx\frac{P(\gamma_e)}{\nu(\gamma_e)}= \frac {m_e c^2
\sigma_T} {3 q_e} \gamma B .
\end{equation}
Note that $P_{\nu,max}$ does not depend on $\gamma_e$, whereas the
position of the peak does.

The above description of $P_\nu$ describes the emitted spectrum when
the electron does not lose a significant fraction of its energy to
radiation. This requires $\gamma_e$ to be less than a critical value
$\gamma_c$, above which cooling by synchrotron radiation is
significant. The critical electron Lorentz factor $\gamma_c$ is
given by the condition $\gamma\gamma_cm_ec^2=P(\gamma_c)t$,
\begin{equation}
\label{cool}
\gamma_c= \frac {6 \pi m_e c} {\sigma_T \gamma B^2 t}= 
\frac {3 m_e} {16 \epsilon_B \sigma_T m_p c} \frac 1 {t \gamma^3 n},
\end{equation}
where $t$ refers to time in the frame of the observer.

Consider now an electron with an initial Lorentz factor
$\gamma_e>\gamma_c$. This electron cools down to $\gamma_c$ in the
time $t$. As it cools, the frequency of the synchrotron emission
varies as $\nu\propto \gamma_e^2$ while the electron energy varies as
$\gamma_e$. It then follows that the spectral power varies as
$\nu^{-1/2}$ over the frequency range $\nu_c<\nu<\nu(\gamma_e)$, where
we have defined $\nu_c\equiv\nu(\gamma_c)$. The net spectrum of
radiation from such an electron then consists of three segments, a low
energy tail for $\nu<\nu_c$ where $P_\nu$ goes as $\nu^{1/3}$, a
power-law segment between $\nu_c$ and $\nu(\gamma_e)$ where
$P_\nu\sim\nu^{-1/2}$, and an exponential cutoff for
$\nu>\nu(\gamma_e)$. The maximum emissivity occurs at $\nu_c$ and is
given by $P_{\nu,max}$.

As described in the previous section, we are interested in a power-law
distribution of electrons. To calculate the net spectrum due to all
the electrons we need to integrate over $\gamma_e$. There are now two
different cases, depending on whether $\gamma_m>\gamma_c$ or
$\gamma_m<\gamma_c$. 

Let the total number of electrons be $N_e$. When $\gamma_m>\gamma_c$,
all the electrons cool down roughly to $\gamma_c$ and the flux at
$\nu_c$ is approximately $N_e P_{\nu,max}$. We call this the case of
{\it fast cooling}. The flux at the observer, $F_\nu$, is given by
\begin{equation}
\label{spectrumfast}
F_\nu=\cases{ ( \nu / \nu_c )^{1/3} F_{\nu,max}, &
$\nu_c>\nu$, \cr ( \nu / \nu_c )^{-1/2} F_{\nu,max}, &
$\nu_m>\nu>\nu_c$, \cr ( \nu_m / \nu_c )^{-1/2} ( \nu / \nu_m)^{-p/2}
F_{\nu,max}, & $\nu>\nu_m$, \cr }
\end{equation}
where $\nu_m \equiv \nu(\gamma_m)$ and
$F_{\nu,max}\equiv N_eP_{\nu,max}/4\pi D^2$ is the observed peak flux
at distance $D$ from the source.

When $\gamma_c>\gamma_m$, only those electrons with
$\gamma_e>\gamma_c$ can cool. We call this {\it slow cooling},
because the electrons with $\gamma_e\sim\gamma_m$, which form the bulk
of the population, do not cool within a time $t$. Integrating over
the electron distribution gives
\begin{equation}
\label{spectrumslow}
F_\nu=\cases{ 
(\nu/\nu_m)^{1/3} F_{\nu,max},
            & $\nu_m>\nu$, \cr
(\nu/\nu_m)^{-(p-1)/2} F_{\nu,max},
            & $\nu_c>\nu>\nu_m$, \cr
\left( \nu_c/\nu_m \right)^{-(p-1)/2} 
\left( \nu/\nu_c \right)^{-p/2} F_{\nu,max},
            & $\nu>\nu_c$. \cr
}
\end{equation}

Typical spectra corresponding to fast and slow cooling are shown in
Figures 1a and 1b. In addition to the various power-law regimes
described above, self-absorption causes a steep cutoff of the
spectrum, either as $\nu^2$ or $\nu^{5/2}$, at low frequencies (Katz
1994, Waxman 1997b, Katz and Piran 1997). For completeness, we show
this regime in Figure 1, but we shall ignore it for the rest of this
Letter since self-absorption does not affect either the optical or
X-ray radiation in which we are interested.

\begin{figure}
\begin{center}
\epsscale{.75}
\plotone{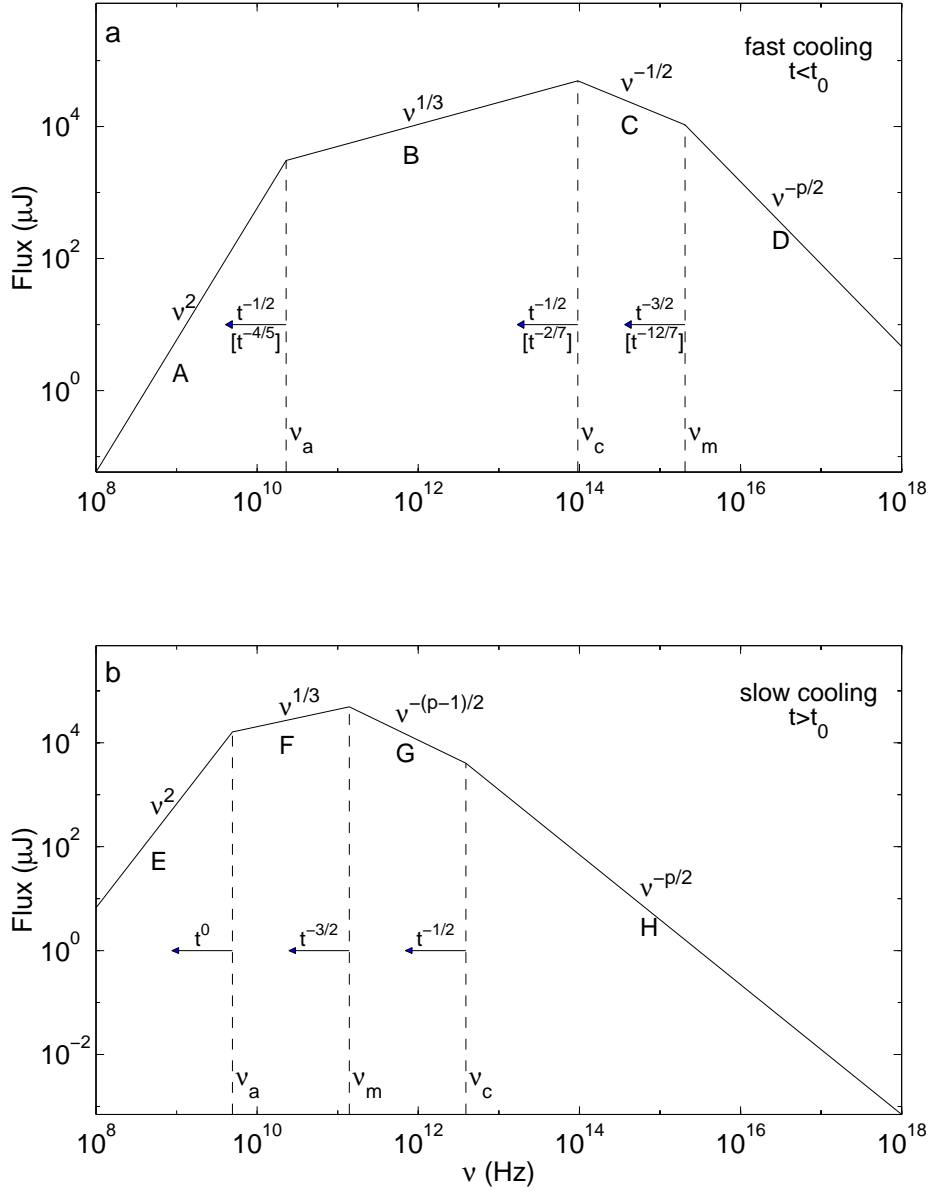}
\caption{ 
  Synchrotron spectrum of a relativistic shock
  with a power-law distribution of electrons. (a) The case of fast
  cooling, which is expected at early times ($t<t_0$) in a
  $\gamma$-ray burst afterglow. The spectrum consists of four
  segments, identified as A, B, C, D.  Self-absorption is important
  below $\nu_a$. The frequencies, $\nu_m$, $\nu_c$, $\nu_a$, decrease
  with time as indicated; the scalings above the arrows correspond to
  an adiabatic evolution, and the scalings below, in square brackets,
  to a fully radiative evolution. (b) The case of slow cooling, which
  is expected at late times ($t>t_0$). The evolution is always
  adiabatic. The four segments are identified as E, F, G, H. 
}
\end{center}
\end{figure}

\section{Hydrodynamical Evolution and Light Curves} 

The instantaneous spectra described in the previous section do not
depend on the hydrodynamical evolution of the shock. The only
assumption made there is that the shock properties are fairly constant
over a time scale comparable to the observation time $t$. The {\it
light curves} at a given frequency, however, depend on the temporal
evolution of various quantities, such as the break frequencies $\nu_m$
and $\nu_c$ and the peak power $N_e P_{\nu,max}$. These depend, in
turn, on how $\gamma$ and $N_e$ scale as a function of $t$.

We limit the discussion here to the case of a spherical shock of
radius $R(t)$ propagating into a constant surrounding density $n$.
Clearly, the total number of swept-up electrons in the post-shock
fluid is $N_e=4\pi R^3 n/3$. We consider two extreme limits for the
hydrodynamical evolution of the shock: either fully radiative or fully
adiabatic. The radiative solution assumes that all the internal energy
created in the shock is radiated. This requires two conditions to be
satisfied: (1) the fraction of the energy going into the electrons
must be large, i.e. $\epsilon_e\to1$, and (2) we must be in the regime
of fast cooling, $\gamma_c<\gamma_m$. If either of these conditions
is not satisfied, i.e. if $\epsilon_e\ll1$ or $\gamma_c\gg\gamma_m$,
then we have adiabatic evolution. 

In the adiabatic case, the energy $E$ of the spherical shock is
constant and is given by $E=16\pi \gamma^2 R^3 n m_p c^2/17$
(Blandford \& McKee 1976, Sari 1997). In the radiative case, the
energy varies as $E\propto\gamma$, where $\gamma \cong (R/L)^{-3}$.
Here $L=[17M/(16\pi m_pn)]^{1/3}$ (Blandford \& McKee 1976, Vietri
1996, Katz \& Piran 1997) is the radius at which the mass swept up
from the external medium equals the initial mass $M$ of the ejecta (We
used $17/16$ instead of $3/4$ in order to be compatible with the
adiabatic expression and to enable a smooth transition between the
two); we write $M$ in terms of the initial energy of the explosion via
$M=E/\gamma_0c^2$, where $\gamma_0$ is the initial Lorentz factor of
the ejecta.

In both the adiabatic and radiative cases, there is a simple relation
connecting the radius of the shock $R$, the fluid Lorentz factor
$\gamma$, and the observed time $t$: $t=R/c_t \gamma^2c$, where the
numerical value of $c_t$ lies between $\sim3$ and $\sim 7$ depending
on the details of the hydrodynamical evolution and the spectrum (Sari
1997a, Waxman 1997c, Sari 1997b, Panaitescu \& M\'esz\'aros 1997). For
simplicity we use $t \cong R/4\gamma^2c$ for all cases. We then have
the following hydrodynamical evolution equations,
\begin{equation}
R(t) \cong \cases {
(17Et/4\pi m_p n c)^{1/4}, & adiabatic, \cr
(4ct/L)^{1/7} L, & radiative,}
\end{equation}
\begin{equation}
\gamma(t) \cong \cases {
(17E/1024\pi n m_p c^5 t^3)^{1/8}, & adiabatic, \cr
(4ct/L)^{-3/7}, & radiative.}
\end{equation}
Using these scalings and the results of the previous section, we can
calculate the variation with time of all the relevant quantities. For
the adiabatic case, we find
\begin{eqnarray}
\label{abreaks}
\nu_c                  & = & 2.7 \times 10^{12}  \epsilon_B^{-3/2}
                             E_{52}^{-1/2} n_1^{-1} t_d^{-1/2}
                             {\rm \ Hz}, \cr
\nu_m              & = & 5.7 \times 10^{14} \epsilon_B^{1/2} \epsilon_e^2
                             E_{52}^{1/2} t_d^{-3/2} {\rm \ Hz}, \cr
F_{\nu,max}& = & 1.1 \times 10^5  \epsilon_B^{1/2} E_{52} 
n_1^{1/2} D_{28}^{-2} \ \mu {\rm J},
\end{eqnarray}
where $t_d$ is the time in days, $E_{52}=E/10^{52}$ ergs, $n_1$ is $n$
in units of ${\rm cm^{-3}}$ and $D_{28}=D/10^{28}$ cm.
In the case of a fully radiative evolution, the results are
\begin{eqnarray}
\label{rbreaks}
\nu_c                  & = & 1.3 \times 10^{13} \epsilon_B^{-3/2} 
                             E_{52}^{-4/7} \gamma_2^{4/7} 
                              n_1^{-13/14} t_d^{-2/7} {\rm \ Hz}, \cr
\nu_m              & = & 1.2 \times 10^{14} \epsilon_B^{1/2} \epsilon_e^2 
                             E_{52}^{4/7} \gamma_2^{-4/7} n_1^{-1/14}
                             t_d^{-12/7} {\rm \ Hz}, \cr
F_{\nu,max}& = & 4.5 \times 10^3 \epsilon_B^{1/2} E_{52}^{8/7}\gamma_2^{-8/7} 
                            n_1^{5/14} D_{28}^{-2}
                             t_d^{-3/7} \  \mu {\rm J},
\end{eqnarray}
where we have scaled the initial Lorentz factor of the ejecta by a
factor of 100: $\gamma_2\equiv\gamma_0/100$.

The spectra presented in Figure 1 show the positions of $\nu_c$ and
$\nu_m$ for typical parameters. Note that, in both the adiabatic and
radiative cases, $\nu_c$ decreases more slowly with time than $\nu_m$.
Therefore, at sufficiently early times we have $\nu_c<\nu_m$,
i.e. fast cooling, while at late times we have $\nu_c>\nu_m$, i.e.,
slow cooling. The transition between the two occurs when
$\nu_c=\nu_m$.  This corresponds to the time
\begin{equation}
\label{tfc}
t_0=\cases{ 
210 \epsilon_B^2 \epsilon_e^2 E_{52} n_1 ~{\rm days},
       & adiabatic, \cr
4.6 \epsilon_B^{7/5} \epsilon_e^{7/5} E_{52}^{4/5}\gamma_2^{-4/5}
n_1^{3/5} ~{\rm days},
       & radiative.}
\end{equation}
At $t=t_0$, the spectrum changes from fast cooling (Fig. 1a) to slow
cooling (Fig. 1b). In addition, if $\epsilon_e \to 1$, the
hydrodynamical evolution changes from radiative to adiabatic. However,
if $\epsilon_e \ll 1$, the evolution remains adiabatic throughout.

Once we know how the break frequencies, $\nu_c$, $\nu_m$, and the peak
flux $F_{\nu,max}$ vary with time, we can calculate the light
curve. Consider a fixed frequency $\nu=10^{15}\nu_{15}$ Hz. From the
first two equations in (\ref{abreaks}) and (\ref{rbreaks}) we see that
there are two critical times, $t_c$ and $t_m$, when the break
frequencies, $\nu_c$ and $\nu_m$, cross the observed frequency $\nu$:
\begin{equation}
t_c=\cases{ 
7.3 \times 10^{-6} \epsilon_B^{-3} E_{52}^{-1} n_1^{-2} 
\nu_{15}^{-2} ~{\rm days},
         & adiabatic, \cr
2.7 \times 10^{-7} \epsilon_B^{-21/4} E_{52}^{-2} \gamma_2^{2}
n_1^{-13/4} \nu_{15}^{-7/2} ~{\rm days}, 
         & radiative,}
\end{equation}
\begin{equation}
\label{tmin}
t_m=\cases{
0.69 \epsilon_B^{1/3} \epsilon_e^{4/3} E_{52}^{1/3}
\nu_{15}^{-2/3} ~{\rm days},
         & adiabatic, \cr
0.29 \epsilon_B^{7/24} \epsilon_e^{7/6} E_{52}^{1/3} \gamma_2^{-1/3}
\nu_{15}^{-7/12} n_1^{-1/24} ~{\rm days},
         & radiative.}
\end{equation}

It is easily seen that there are only two possible orderings of the
three critical times, $t_c$, $t_m$, $t_0$, namely $t_0>t_m>t_c$ and
$t_0<t_m<t_c$.
Let us define the critical frequency, $\nu_0=\nu_c(t_0)=\nu_m(t_0)$,
\begin{equation}
\label{nu0}
\nu_0=\cases{ 1.8 \times 10^{11} \epsilon_B^{-5/2} \epsilon_e^{-1}
E_{52}^{-1} n_1^{-3/2} \ {\rm Hz}, & adiabatic, \cr 8.5 \times 10^{12}
\epsilon_B^{-19/10} \epsilon_e^{-2/5} E_{52}^{-4/5}\gamma_2^{4/5}
n_1^{-11/10} \ {\rm Hz}, & radiative.}
\end{equation}
When $\nu>\nu_0$, the ordering $t_0>t_m>t_c$ applies and we refer to
the corresponding light curve as the {\it high frequency light curve}.
Similarly, when $\nu<\nu_0$, we have $t_0<t_m<t_c$, and we obtain the
{\it low frequency light curve}.

\begin{figure}
\begin{center}
\epsscale{.75}
\plotone{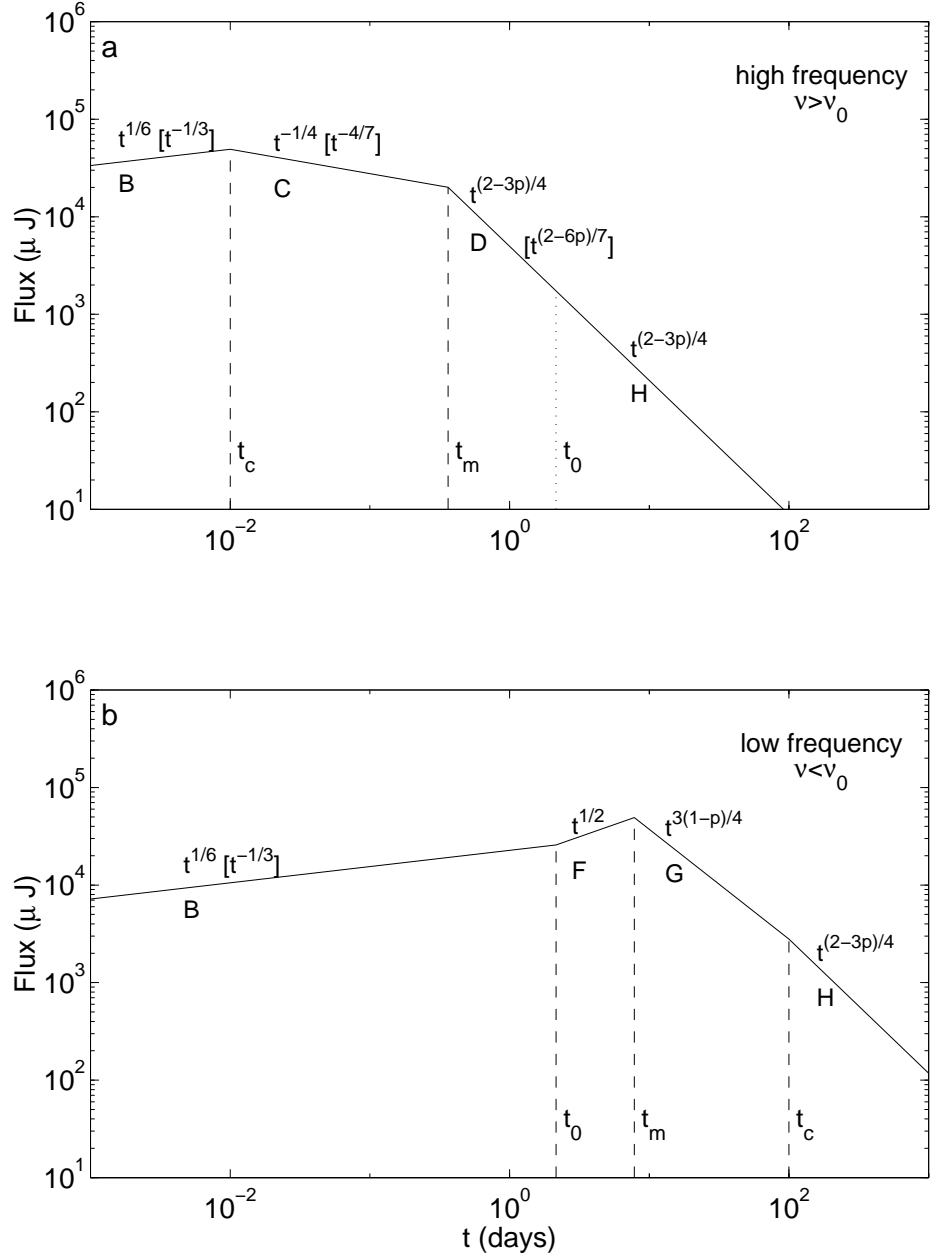}
\caption{ 
  Light curve due to synchrotron radiation from a
  spherical relativistic shock, ignoring the effect of
  self-absorption.  (a) The high frequency case ($\nu>\nu_0$). The
  light curve has four segments, separated by the critical times,
  $t_c$, $t_m$, $t_0$.  The labels, B, C, D, H, indicate the
  correspondence with spectral segments in Fig. 1. The observed flux
  varies with time as indicated; the scalings within square brackets
  are for radiative evolution (which is restricted to $t<t_0$) and the
  other scalings are for adiabatic evolution. (b) The low frequency
  case ($\nu<\nu_0$).  
}
\end{center}
\end{figure}

Figure 2a shows a typical high frequency light curve. At early times
the electrons cool fast and $\nu$ is less than both $\nu_m$ and
$\nu_c$. Ignoring self absorption, the situation corresponds to
segment B in Figure 1, and the flux varies as $F_\nu\sim
F_{\nu,max}(\nu/\nu_c)^{1/3}$.  If the evolution is adiabatic,
$F_{\nu,max}$ is constant, and $F_\nu\sim t^{1/6}$. In the radiative
case, $F_{\nu,max}\sim t_d^{-3/7}$ and $F_\nu\sim t^{-1/3}$. The
scalings in the other segments, which correspond to C, D, H in Fig. 1,
can be derived in a similar fashion and are shown in Fig. 2a.

Figure 2b shows the low frequency light curve, corresponding to
$\nu<\nu_0$. In this case, there are four phases in the light curve,
corresponding to segments B, F, G and H. The time dependences of the
flux are indicated on the plot for both the adiabatic and radiative
cases.

We conclude with two comments. First, during radiative evolution the
energy in the shock decreases with time, the energy to be substituted
in the radiative scalings is the initial energy. When a radiative
shock switches to adiabatic evolution at time $t=t_0$, it is necessary
to use the reduced energy to calculate the subsequent adiabatic
evolution. The final energy $E_{f,52}$ which one should use in the
adiabatic regime is related to the initial $E_{i,52}$ of the fireball
by
\begin{equation}
\label{finalE}
E_{f,52}=0.022\epsilon_B^{-3/5}\epsilon_e^{-3/5}
E_{i,52}^{4/5}\gamma_2^{-4/5}n_1^{-2/5}.
\end{equation}
Second, if during the phase of fast cooling ($t<t_0$) $\epsilon_e$ is
somewhat less than unity, then only a fraction of the shock energy is
lost to radiation. The scalings will be intermediate between the two
limits of fully radiative and fully adiabatic discussed here.

\section{Discussion}
The main results of this Letter are summarized in Fig. 1 and Fig. 2,
along with the scalings given in equations
(\ref{abreaks})--(\ref{finalE}).

It is well-known that the flux at the peak of the synchrotron spectrum
is independent of time in the slow cooling limit for adiabatic
hydrodynamic evolution (M\'esz\'aros \& Rees 1996). We have shown in
this Letter that the peak flux is constant even in the fast cooling
limit if the evolution is adiabatic. The fast cooling stage has
generally not been treated by other authors. We show that the
position of the peak of the spectrum varies as $\nu_c\propto t^{-1/2}$
during fast cooling compared to $\nu_m\propto t^{-3/2}$ in slow
cooling (Fig. 1), and this is one way of distinguishing between the
two cases. We have also derived the scalings for a fully radiative
evolution, where all the shock energy is radiated efficiently. This
regime again has rarely been discussed in the literature. We find
that the peak flux decreases with time as $F_{\nu,max}\propto
t^{-3/7}$ and the position of the peak varies as $\nu_c \propto
t^{-2/7}$. (These results differ from those given in Katz \& Piran
1997, who considered the flux at $\nu_m$ instead of the peak flux,
which is at $\nu_c$.)

Even within the adiabatic case, we find that there are two possible
slopes for the decaying part of the light curve. Writing the flux as
$F_\nu\sim t^{-\alpha}$, the two cases give $\alpha=3p/4-3/4$ and
$\alpha=3p/4-1/2$. If the physics of particle acceleration in
relativistic shocks is universal in the sense that the power law index
$p$ of the electron distribution is always the same, and if the
evolution is adiabatic, then we expect always to observe one of these
two values of $\alpha$, which differ by $1/4$. Indeed, some X-ray
afterglows appear to decay with $\alpha\cong 1.4$ while the optical
and X-ray afterglows of GRB 970228 and GRB970508 had $\alpha \cong
1.2$ (Yoshida et al. 1997, Sokolov et. al. 1997). The difference
between the two values is consistent with 1/4. The corresponding value
of $p$ is $\sim2.5-2.6$, which is a reasonable energy index for shock
acceleration.
If future observations of $\gamma$-ray burst afterglows always find
decays with either $\alpha=1.4$ or $\alpha=1.2$, it will be a strong
confirmation of the shock model and the adiabatic assumption. The
characteristic values of $\alpha$ are different for radiative
evolution.

In addition to the decay of the light curve with time, we can also
consider the spectral index $\beta$, defined by $F_\nu\sim
\nu^{-\beta}$. The two values of $\alpha$ given above for adiabatic
evolution correspond to $\beta=(p-1)/2$ and $\beta=p/2$,
respectively. Thus, the relation between $\alpha$ and $\beta$ in an
adiabatic fireball is either $\alpha=3\beta/2$ or
$\alpha=3\beta/2+1/2$. Previous studies (Rees \& M\'esz\'aros 1996,
Waxman 1997a) have considered the first possibility. However, we note
that, for the standard choice of parameters, namely $n\sim1~{\rm
cm^{-3}}$ (a standard interstellar medium), $\epsilon_e,~\epsilon_B >
0.1$ (rough equipartition of energy between electrons and magnetic
fields), the second relation holds in both the optical and X-ray bands
during much of the decay. Indeed, this relation is more compatible
with detailed observations of GRB 970508 (Sokolov et. al. 1997).

Finally, we note that in none of the cases considered does the flux
rise more steeply than $t^{1/2}$. This is a potential problem since
GRB970508 displayed a sharp rise in the optical flux just before its
peak at two days (Sokolov et. al. 1997).

\acknowledgments This work was supported by NASA Grant NAG5-3516, and
a US-Israel BSF Grant 95-328. Re'em Sari thanks The Clor Foundation
for support.


\begin{references}
\reference{BlandfordMcKee} Blandford, R. D. \& McKee, C. F. 1976, 
Phys. of Fluids, 19, 1130
\reference{Costa1} Costa, E. et al. 1997a, IAU Circular No. 6572
\reference{Frail} Frail, D. A., Kulkarni, S. R., Nicastro, L.,
Feroci, M., \& Taylor, G. B. 1997, Nature, in press.
\reference{Goodman} Goodman, J. 1997, New Astronomy 2(5): 449-460.
\reference{Groot} Groot, P. J. et al., 1997 IAU Circular No. 6584
\reference{KatzPiran} Katz \& Piran 1997, \apj \ December 1, astro-ph/9706141.
\reference{MeszarosRees97} M\'esz\'aros, P. \& Rees, M. 1997, \apj, 476, 232
\reference{PanaitescuMeszaros97a} Panaitescu A. \& M\'esz\'aros, P., 1997,
  submitted to \apj, astro-ph/9709284
\reference{Sari97a} Sari, R. 1997a, \apj L, 489, L37.
\reference{Sari97b} Sari, R. 1997b, submitted to \apj L,
astro-ph/9709300.
\reference{SNP} Sari, R., Narayan, R. \& Piran, T. 1996, \apj, 473, 204.
\reference{Sokolov} Sokolov et. al. 1997, astro-ph/9709093.
\reference{Vietry} Vietri, M. 1997, \apj L, 478, L9.
\reference{WijersReesMeszaros} Wijers, R. A. M. P., Rees M. \&
M\'esz\'aros, P. 1997, MNRAS, 288, L51.
\reference{Waxman97a} Waxman, E. 1997a, \apj, 485, L5.
\reference{Waxman97b} Waxman, E. 1997b, Nature, submitted, astro-ph/9705229
\reference{Waxman97c} Waxman, E. 1997c, \apj, submitted,
astro-ph/9705229
\reference{Yoshida} Yoshida et al., 1997, to appear in proceeding of the 4th
Huntsville symposium.
\end{references}
\end{document}